\documentclass[preprint,12pt]{elsarticle}
\usepackage{color,soul}
\usepackage{graphicx}
\usepackage{algpseudocode}
\usepackage{setspace}
\usepackage{booktabs}
\usepackage[font=small,skip=0pt]{caption}
\usepackage{amsthm}
\usepackage{hyperref}
\usepackage{amssymb}
\usepackage{commath}
\usepackage[ruled, vlined, linesnumbered]{algorithm2e}
\usepackage{bm}

\makeatletter
\newcommand{\LeftEqNo}{\let\veqno\@@leqno}
\makeatother

\journal{Journal of Biotechnology} 	
\begin{document}
\begin{frontmatter}

\title{CAMIRADA: Cancer microRNA association discovery algorithm, a case study on breast cancer}

\author[add1]{Sepideh Shamsizadeh\corref{cor1}}
\ead{s.shamsizadeh@ut.ac.ir}

\author[add1]{Sama Goliae}
\ead{sgoliaei@ut.ac.ir}

\author[add2]{Zahra Razaghi moghadam}
\ead{mona.razaghi@gmail.com}

\cortext[cor1]{Corresponding author}
\address[add1]{Faculty of New Sciences and Technologies, University of Tehran, Tehran, Iran}
\address[add2]{Max Planck Institute of Molecular Plant Physiology, Posdam, German}

\begin{abstract}In recent studies, non-coding protein RNAs have been identified as microRNA that can be used as biomarkers for early diagnosis and treatment of cancer, that decrease mortality in cancer. A microRNA may target hundreds or thousands of genes and a gene may regulate several microRNAs, so determining which microRNA is associated with which cancer is a big challenge. Many computational methods have been performed to detect micoRNAs association with cancer, but more effort is needed with higher accuracy. Increasing research has shown that relationship between microRNAs and TFs play a significant role in diagnosis of cancer. Therefore, we developed a new computational framework (CAMIRADA) to identify cancer-related microRNA based on the relationship between microRNAs and disease genes (DG) in the protein network, the functional relationships between microRNAs and Transcription Factors (TF) on the co-expression network, and the relationship between microRNAs and the Differential Expression Gene (DEG) on co-expression network. The CAMIRADA was applied to assess breast cancer data from two HMDD and miR2Disease databases. In this study, the AUC for the 65 microRNA of the top of the list was 0.95, which was more accurate than the similar methods used to detect microRNA associated with the cancer artery.
\end{abstract}

\begin{keyword}
microRNAs \sep Transcription factors \sep Differentially expressed gene \sep Diseases gene \sep Co-expression network
\end{keyword}

\end{frontmatter}

\section{Introduction} \label{I}

According to a study,breast cancer is one of the most common causes of women's deaths in cancer \citep{yang2015crucial}. The molecular traits of the primary tumors play an important role in timely diagnosis and treatment of the next stages of breast cancer. Therefore, we need methods that can detect cancer at an early stage in order to help clinicians and patients to be treated with biological markers \citep{yang2015crucial}.
In recent years, many studies have been conducted to clarify the mechanisms that make cancer progress and develop.

Although many genes that cause cancer and stop it are known by researchers, it is still necessary to identify cancer pathways. Therefore, one of the most important biological targets of cancer is the detection of genes related to cancers \citep{farahmand2016gta,razaghi2016hybridranker}. 

MicroRNAs are short non-coding RNAs with an approximate length of 22 nucleotides that are involved in post-transcriptional regulation and that are major regulators of gene expression \citep{guruceaga2014functional}.In 2002, the first link between cancer and microRNAs was identified \citep{adali2012analysis}. MicroRNAs target about 60\% of human genes that be involved in a wide range of biological processes and disease including cell division, proliferation, differentiation, and apoptosis \citep{le2015network}. More than 50\% of human microRNAs are located in cancer-associated genomic regions. MicroRNAs could play roles as oncogenes or tumor suppressor genes. Dysregulation of microRNA expression has been shown to have impacts on human diseases \citep{le2015network}. 

However, identifying the related microRNAs with existing experimental tests may be difficult and time consuming. In addition, many researchers are also faced with limited knowledge about microRNA. Therefore, a number of computational approaches have been recently developed to identify microRNAs associated with the disease. Statistical methods, machine learning, and network-based methods are some of the methods proposed to predict cancer-related microRNAs.

Zhao et al. developed a framework for obtaining the relationship between cancer and microRNAs through their target gene expression profiles without requiring neither microRNA expression data or the matched gene and microRNA expression data.

For each microRNA, target genes that are likely to be related to cancer are determined, and subsequently identified adverse pathways associated with cancer. Then, microRNAs are ranked according to these pathways, because high-ranking microRNAs are more likely to be linked to cancer \citep{zhao2014identifying}.

Xue et al. used the Gaussian mixture models, they identified patterns of gene expression for healthy and patient samples, and then, with Fisher's exact test, deduced the regulator relationship between TF and genes. A minimum description length for the pruning of the network, using the (MDL) principle, the relationships between TFs and microRNAs have been achieved \citep{xue2017computational}.

Tseng et al. provided a method to identify active oncomirs and their potential functions in gastric cancer progression. The microRNA and mRNA's expression profiles with the human protein interaction network (PIN) are integrated to show microRNA-regulated PIN in specific biological conditions. The microRNAs' potential functions were identified by functional enrichment analysis and the activities of microRNA-regulated PINs were evaluated by the co-expression of protein-protein interactions (PPIs) \citep{tseng2011integrative}.

Le constructed microRNA functional similarity networks based on shared targets of microRNAs, and then it integrated them with a microRNA functional synergistic network. After analyzing topological properties of these networks, it introduced five network-based ranking methods (RWR, PRINCE (Prioritization and Complex Elucidation), which was proposed for disease gene prediction; PageRank with Priors (PRP) and K-Step Markov (KSM), which were used for studying web networks; and a neighborhood-based algorithm.) in prioritizing candidate microRNAs to predict novel disease-related microRNAs based on the constructed microRNA functional similarity networks \citep{le2015network}.

In this study, we presented CAMIRADA framework for the identification of microRNA cancer associations by taking advantage of the useful links between microRNA targets and disease genes in protein-protein networks (PPIs), the functional connections between microRNA targets and Transcription Factors(TF) in co-expression network and the functional relationships between microRNA targets and Differentially Expressed Gene(DEG) in co-expression network.

\section{Methods}\label{II}
\subsection{Data Sets}
We firstly collected the microRNA target genes predicted with different tools, including PicTar \citep{krek2005combinatorial}, miRanda (version 3.0) \citep{john2004human}, TargetScan (release 6.2) \citep{lewis2005conserved}, miRBase (version 16) \citep{griffiths2006mirbase} and mirTarbase \citep{pinero2016disgenet}. Specifically, for a microRNA, we consider target genes to be obtained at least by three tools. Note that some microRNAs are not considered if, after filtering, its target genes are less than three. In total, we obtained 42832 targeting pairs that involved 825 microRNAs and 8334 target genes.

The disease-gene association data were obtained from DisGeNET \citep{abbott2014candidate} and The Candidate Cancer Gene Database (CCGD) \citep{yang2012chipbase}, which involved 375 genes related to breast cancer. The Transcription factors(TF) were retrieved from ChIPBase \citep{chou2015mirtarbase}, that included 13890 genes. The names of all these genes mapped to Entrez gene id.

\subsection{Human Protein-Protein Interaction (PPI) Data And Co-Expression Network}
PPI data for humans are derived from the Human Protein Reference Database (HPRD Release 9), which contains human-protein annotations based on empirical evidence from published reports \citep{shi2013walking}. We changed the name of the gene to the Entrez gene id, and then we obtained the maximum components of the entire network, which contains 9028 genes and 35865 interactions. It is noteworthy that PPI data in HPRD are the most common protein isoforms, mainly due to lack of experimental data \citep{lewis2005conserved}. 

We downloaded GSE31192 \citep{harvell2013genomic} data from the GEO \citep{barrett2012ncbi} database to build a co-expression network and obtain DEGs. The RNA extracted from breast tissue with microarrays Affymetrix Human Genome U133 Plus 2.0  has been investigated. These gene expression data are pre-processed with the Robust Multichip Average algorithm(RMA). 

Then we find out the Differentially Expressed Genes (DEGs) between normal and abnormal samples with samr package in R. We selected genes, which their $p-value$ were under 0.05 and their $nfold$ was 0.1 . The co-expression network was constructed using the WGCNA \citep{langfelder2008wgcna} package in R software. To choose an appropriate cutoff to include a percentage of the highest correlations, the approximate scale-free topology criterion was applied. For determining the optimal parameter, the function pickSoftThresholdthe in WGCNA package was used. After selecting the best threshold (0.85), adjacency network for co-expression network was calculated.

 The co-expression network included 21179 genes, so for sampling, we considered DEGs and all targets of microRNAs as seed, we mapped them onto the co-expression network, and after that, we applied RWR (Random Walk With Restart)  algorithm for ranking the genes of network \citep{shi2013walking}. Then we selected top 10000 genes and made the network base on their relationships.

\subsection{CAMIRADA algorithm}
\subsubsection{Disease Genes-microRNA}
We mapped targets of a microRNA and disease genes onto PPI network. After that, we considered disease genes as seeds and applied RWR \citep{shi2013walking} to rank genes of PPI network. Then by using gene set enrichment analysis (GSEA) we derived the ranked gene list, we used ES1 (enrichment score) using the following formula \citep{shi2013walking}.

\begingroup\makeatletter\def\f@size{9}\check@mathfonts
	\begin{equation}\label{eq:01}
		\text{ES1} = \max_{1\le i \le n} \left( \sum_{\substack{g_j \in TG,\\ j\le i} } \sqrt{\frac{N-n_1}{n_1}} - 		\sum_{\substack{g_j \notin TG,\\ j\le i} } \sqrt{\frac{N-n_1}{n_1}}\right)
	\end{equation}
	\endgroup

Where $N$ is the number of all genes in PPI network and $n_1$ is the number of target genes of one microRNA that we showed with TG.  For each microRNA we calculate ES1. We computed a statistical sum from the beginning of the ranked list of PPI's genes. In this way, by moving down the list if the gene was in TG, we added to statistical sum, and if that gene did not exist in the TG, we reduced from statistical sum. The RWR algorithm with each microRNA target genes as seeds was applied to compute ES2 for the same pairing of microRNA disease referred to above \citep{shi2013walking}.

\begingroup\makeatletter\def\f@size{9}\check@mathfonts
	\begin{equation}\label{eq:02}
		\text{ES2} = \max_{1\le i \le n} \left( \sum_{\substack{g_j \in DG,\\ j\le i} } \sqrt{\frac{N-n_1}{n_1}} - 	\sum_{\substack{g_j \notin DG,\\ j\le i} } \sqrt{\frac{N-n_1}{n_1}}\right)
	\end{equation}
	\endgroup

Which $n_2$ is the number of disease genes. For all pairs of microRNA and disease genes we did the above procedures. For computing ES1 we considered disease genes as seed for RWR algorithm and we considered targets of one microRNA as seed for RWR to compute ES2. Then we calculated ES for each pair of microRNA-disease by using the following formula \citep{shi2013walking}

\begingroup\makeatletter\def\f@size{9}\check@mathfonts
\begin{equation} \label{eq:03}
	\text{ES} = \beta\ \text{ES1} + (1-\beta)\ \text{ES2}
\end{equation}
\endgroup

\subsubsection{DEGs-microRNAs}
We mapped targets of each microRNA and DEGs onto co-expression network. For all pairs of microRNA and DEGs we calculated ES. For computing ES1 we considered DEGs as seed for RWR algorithm and we considered targets of one microRNA as seed for RWR to compute ES2. Then we calculated ES for each pair of microRNA-DEG Where $N$ is the number of all genes in co-expression network and  $n_1$ is the number of target genes of one microRNA.  For each microRNA we calculate ES1  \ref{eq:01}. Which $n_2$  in the ES2  \ref{eq:02} is the number of DEGs.

\begin{figure}[!t]
		\centering
		\includegraphics[width=.9 \textwidth]{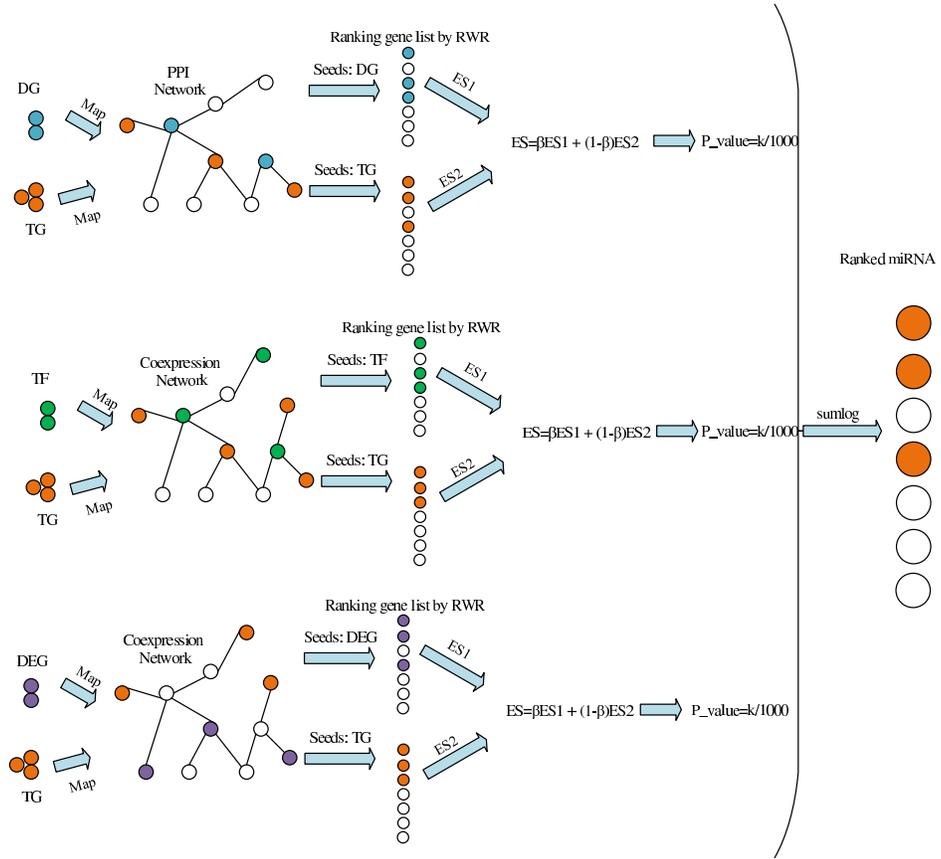}
		\caption{An overview of our framework to predict new diseases associated with microRNA.}\label{fig:1}		
\end{figure}

\subsubsection{TFs-microRNAs}
We did all above producers on TF and targets of microRNAs. First in order, we mapped targets of all microRNAs and TFs onto co-expression network and for each microRNA we separately saved it's targets in a list. For all pairs of microRNA and TFs we calculated ES \ref{eq:03}. For computing ES1 we considered TFs as seed for RWR algorithm and we considered targets of one microRNA as seed for RWR to compute ES2. Then we calculated ES for each pair of microRNA-TF Where $N$ is the number of all genes in co-expression network and $n_1$ is the number of target genes of one microRNA.  For each microRNA we calculate ES1 \ref{eq:01}. Which $n_2$ in the ES2 \ref{eq:02} is the number of TFs. A review of our frame work are showed in Figure \ref{fig:1}.

\subsection{Random Network And P-value}
To measure the importance of the relationship between microRNA and disease, we used the p-value. To calculate the p-value, we first created 1000 random networks of the PPI network and 1000 random networks of the co-expression network. Random networks were made in such a way that the number of input and output edges from each node with the number of input and output edges from the same node in the primary network are equal \citep{shi2013walking}. Of course, it should be noted that both the PPI network and the co-expression network are non-directional networks in this study. The p-value is calculated based on \citep{shi2013walking}.

\section{Results And Performance}\label{III}
To evaluate the performance of our algorithm to identify microRNA-disease associations, we plotted the receiver operating characteristic (ROC) curves and computed the area under the curve (AUC). 

To compute the AUC, we first need to calculate the sensitivity and specificity based on the following formulas, and the sensitivity is defined as the correct detection rate of the positive group on the $y$ axis and the specificity, or the wrong detection rate of the negative category on the $x$ axis. A ROC curve shows a relative compromise between profits and costs \citep{le2015network}.

\begingroup\makeatletter\def\f@size{9}\check@mathfonts
\begin{equation}\label{eq:04}
	1 - \text{specificity} = \frac{FP}{TN+FP}
\end{equation}
\endgroup
\begingroup\makeatletter\def\f@size{9}\check@mathfonts
\begin{equation}\label{eq:05}
	1 - \text{Sensitivity} = \frac{TP}{FN+TP}
\end{equation}
\endgroup

For computing of these we needed to define the set of $TP$، $TN$، $FP$ and $FN$. At first we should define positive set and negative set. For positive set we considered the known cancer related microRNAs were obtained from miR2Disease \citep{jiang2008mir2disease} and HMDD \citep{li2013hmdd} databases.

Nowadays, it is very difficult or impossible to collect data for non-cancerous microRNA \citep{lee2013microrna}. For a negative data set, we chose microRNAs that exhibited the lowest fold change values as negative controls by analyzing the corresponding expression profile of the breast cancer. We also used the same number of negative controls as positive ones. We downloaded microRNA expression profile GSE45666 \citep{lee2013microrna} from GEO. 

We computed three lists of p-value for microRNAs according to our method. We arranged lists of p-value by descending order and we calculated AUC for each lists to identify the importance of each factors (TF, disease and DEG) for determining which microRNA is associated with breast cancer.

\begin{figure}[!t]
		\centering
		\includegraphics[width=.8 \textwidth]{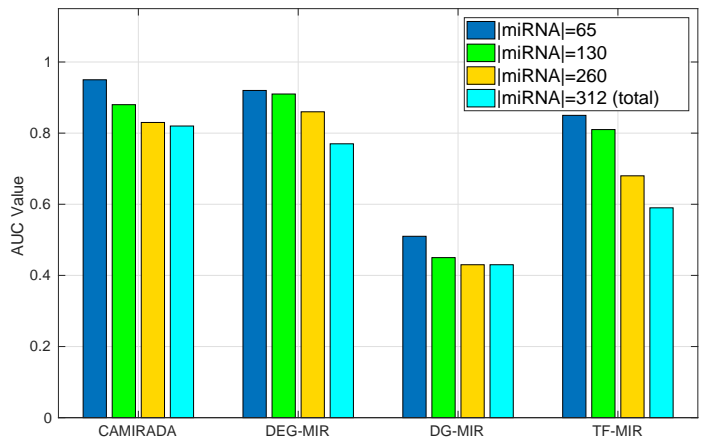}
		\caption{Computing AUC for each three list of p-value for DEG-microRNA, Disease-microRNA and TF-microRNA. And we calculated AUC for CAMIRADA.}\label{fig:2}	
\end{figure}

 As the Figure \ref{fig:2} shows, the lowest AUC is for disease-microRNA and the maximum AUC is for CAMIRADA approach. So, this shows that DEGs and TFs influence on identification of which microRNAs are related to breast cancers. For using the influence of TFs and DEGs on identification of microRNAs, we combined p-vlues with some tests in R.

The tests included Wilkinsonp, sumlog, sumz, logitp, meanp and sump from metap \citep{MetaP} package in R. Then we scored microRNA according to our model. For each p-value as threshold, we computed $TP$, $TN$, $FP$ and $FN$ by using the following definition:
\begin{itemize}
\item TP: The number of microRNAs are in positive set and their scores are below the threshold.
\item FP: The number of microRNAs are in negative set or they are not in positive set and their scores are below the threshold.
\item TN: The number of microRNAs are in negative set or they are not in positive and their scores are above the threshold.
\item FN: The number of microRNAs are in positive set and their scores are above the threshold.
\end{itemize}

\begin{figure}[!t]
		\centering
		\includegraphics[width=.9 \textwidth]{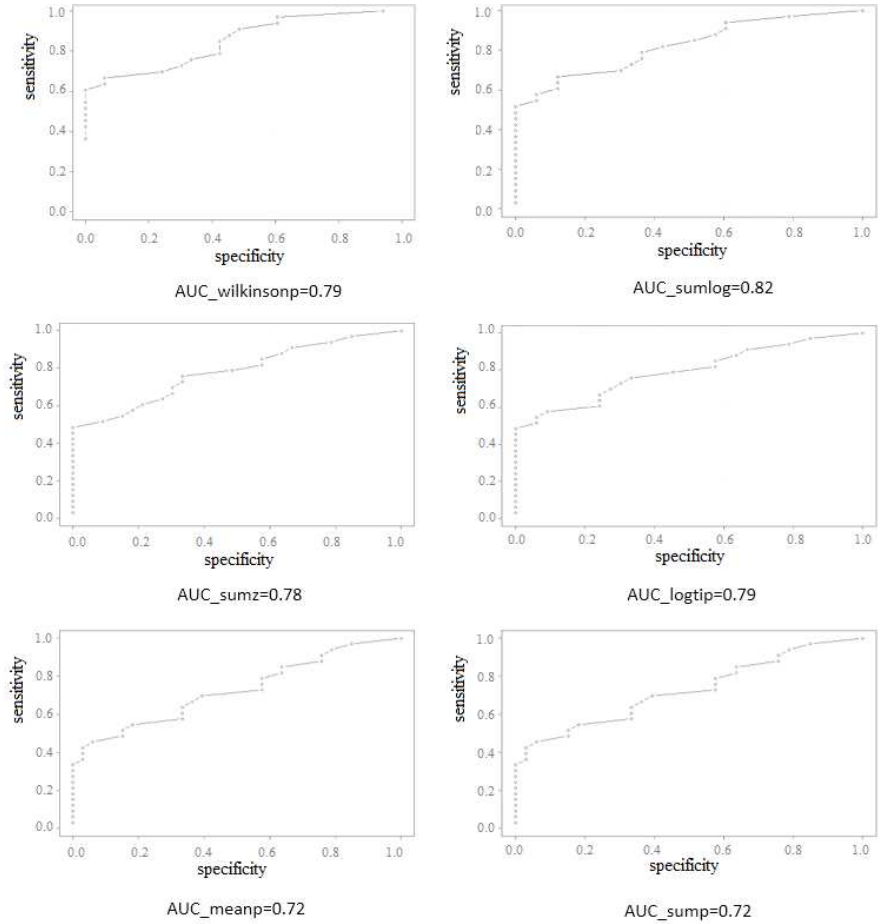}
		\caption{The AUC for all of the p-value combining tests.}\label{fig:3}	
\end{figure}

To select the best test for combining the p-value, we combined the three lists of p-values by all tests and then we calculated AUC for each tests results are shown in Figure \ref{fig:3}. As shown in Figure \ref{fig:3}, the best result is for sumlog test by 82\%, so we choose this test for combining the three lists of p-values.

\begin{figure}[!t]
		\centering
		\includegraphics[width=.8\textwidth]{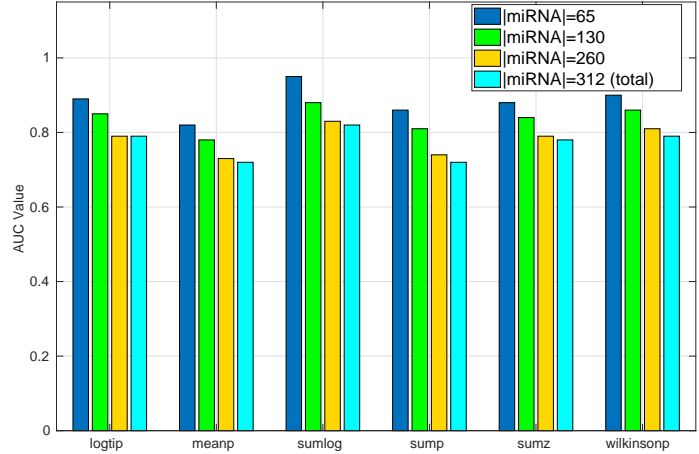}
		\caption{Bar chart for comparing the tests to combine the p-value.}\label{fig:4}	
\end{figure}

For more comparing the tests to combine the p-value we combined the three lists of p-values and we calculated AUC for 65, 130, and 260 number of the microRNA from the top of the ranked list of microRNA and the result of the comparison of tests are shown by bar chart in Figure \ref{fig:4}.

According the results that achieved from tests of p-value combining, sumlog choose to combine p-value. So, we combined three ranked list of p-value with sumlog and then arranged the list of combined p-vlaues in ascending order. The lower the p-value for a microRNA, the greater the probability that this microRNA is associated with the cancer. 

In Table \ref{tab:Table 1}  the 10 top of the list of the ranked microRNA with CAMIRADA and the role of microRNAs in breast cancer are shown.

\begin{table}
	\centering
	\caption{The 10 top of the list of the ranked microRNA.}
	\label{tab:Table 1}
	\begin{tabular}{lll}
		\toprule
		microRNA & p-value & role of microRNA in breast cancer\\
		\midrule
		hsa-miR-93 \citep{shyamasundar2016mir} & 0.000679 & Tumor suppressor \\
		hsa-miR-526b \citep{landman2014role} & 0.000702 & Oncogene \\
		hsa-miR-17 \citep{hossain2006mir} & 0.000951 & Oncogene \\
		hsa-miR-20b \citep{li2012differential} & 0.00125 & Oncogene \\
		hsa-miR-199a \citep{chen2016mir} & 0.001486 & Tumor suppressor \\
		hsa-miR-20a \citep{li2015epigenetic} & 0.001529 & Oncogene \\
		hsa-miR-149 \citep{fang2016down} & 0.0020011 & Oncogene \\
		hsa-miR-199b \citep{lehmann2010identification} & 0.00265 & Biomarker \\
		hsa-miR-519d [38] & 0.002863 & Oncogene \\
		hsa-miR-502 \citep{cheng2012microrna} & 0.036284 & Tumor suppressor\\
		\bottomrule
	\end{tabular}
\end{table}

\section{Conclusion}\label{IV}
The identification of novel cancer-associated microRNAs is important for diagnosing cancer in primary steps for treatment the cancer. In this study, we presented a new approach (CAMIRADA) to identify microRNA related to breast cancer. CAMIRADA find the relationship between microRNA and cancer and it ranked microRNA base on their relation with cancer, by using the microRNA-disease genes associated, the DEGs-microRNA related, and the TFs-microRNAs relation. To evaluated CAMIRADA algorithm we calculated AUC which was 0.95 for 65 microRNAs in top of the ranked list of microRNA. CAMIRADA showed that TFs and DEGs have important roles for identification of which microRNA is related to cancer, so for showing the influence of them we combined three ranked lists of p-values by sumlog test.

By using CAMIRADA, we predicted eight novel microRNAs including hsa-miR-93, hsa-miR-526b, hsa-miR-20b, hsa-miR-199a, hsa-miR-6884, hsa-miR-199b, and hsa-miR-519d associated to breast cancer which are not yet recorded in the disease-microRNA association HMDD and miR2Disease. For future studies, our methods can be used to identify which microRNA are related which cancer, we can make a network for each microRNA and associated cancers, then analysis the network and extract information to identify related between cancers.

\section*{References}
\bibliographystyle{unsrtnat}
\bibliography{Ref}

\begin{thebibliography}{34}
\providecommand{\natexlab}[1]{#1}
\providecommand{\url}[1]{\texttt{#1}}
\expandafter\ifx\csname urlstyle\endcsname\relax
  \providecommand{\doi}[1]{doi: #1}\else
  \providecommand{\doi}{doi: \begingroup \urlstyle{rm}\Url}\fi

\bibitem[Yang et~al.(2015)Yang, Xing, Liang, Hu, Xu, and Chen]{yang2015crucial}
Yang Yang, Yiqiao Xing, Chaoqun Liang, Liya Hu, Fei Xu, and Yuan Chen.
\newblock Crucial micrornas and genes of human primary breast cancer explored
  by microrna-mrna integrated analysis.
\newblock \emph{Tumor Biology}, 36\penalty0 (7):\penalty0 5571--5579, 2015.

\bibitem[Farahmand et~al.(2016)Farahmand, Goliaei, Ansari-Pour, and
  Razaghi-Moghadam]{farahmand2016gta}
S~Farahmand, S~Goliaei, N~Ansari-Pour, and Z~Razaghi-Moghadam.
\newblock Gta: a game theoretic approach to identifying cancer subnetwork
  markers.
\newblock \emph{Molecular BioSystems}, 12\penalty0 (3):\penalty0 818--825,
  2016.

\bibitem[Razaghi-Moghadam et~al.(2016)Razaghi-Moghadam, Abdollahi, Goliaei, and
  Ebrahimi]{razaghi2016hybridranker}
Zahra Razaghi-Moghadam, Razieh Abdollahi, Sama Goliaei, and Morteza Ebrahimi.
\newblock Hybridranker: Integrating network topology and biomedical knowledge
  to prioritize cancer candidate genes.
\newblock \emph{Journal of biomedical informatics}, 64:\penalty0 139--146,
  2016.

\bibitem[Guruceaga and Segura(2014)]{guruceaga2014functional}
Elizabeth Guruceaga and Victor Segura.
\newblock Functional interpretation of microrna--mrna association in biological
  systems using r.
\newblock \emph{Computers in biology and medicine}, 44:\penalty0 124--131,
  2014.

\bibitem[Adal{\i} and {\c{S}}ekero{\u{g}}lu(2012)]{adali2012analysis}
Terin Adal{\i} and Boran {\c{S}}ekero{\u{g}}lu.
\newblock Analysis of micrornas by neural network for early detection of
  cancer.
\newblock \emph{Procedia Technology}, 1:\penalty0 449--452, 2012.

\bibitem[Le(2015)]{le2015network}
Duc-Hau Le.
\newblock Network-based ranking methods for prediction of novel disease
  associated micrornas.
\newblock \emph{Computational Biology and Chemistry}, 58:\penalty0 139--148,
  2015.

\bibitem[Zhao et~al.(2014)Zhao, Liu, Zhu, He, Duval, Richer, Huang, Jiang, Hao,
  and Chen]{zhao2014identifying}
Xing-Ming Zhao, Ke-Qin Liu, Guanghui Zhu, Feng He, B{\'e}atrice Duval,
  Jean-Michel Richer, De-Shuang Huang, Chang-Jun Jiang, Jin-Kao Hao, and Luonan
  Chen.
\newblock Identifying cancer-related micrornas based on gene expression data.
\newblock \emph{Bioinformatics}, 31\penalty0 (8):\penalty0 1226--1234, 2014.

\bibitem[Xue et~al.(2017)Xue, Liu, Zhang, Chang, Liu, Du, Yang, and
  Wang]{xue2017computational}
Mengzhu Xue, Haiyue Liu, Liwen Zhang, Hongyuan Chang, Yuwei Liu, Shaowei Du,
  Yingqun Yang, and Peng Wang.
\newblock Computational identification of mutually exclusive transcriptional
  drivers dysregulating metastatic micrornas in prostate cancer.
\newblock \emph{Nature communications}, 8:\penalty0 14917, 2017.

\bibitem[Tseng et~al.(2011)Tseng, Lin, Chen, Huang, and
  Juan]{tseng2011integrative}
Chien-Wei Tseng, Chen-Ching Lin, Chiung-Nien Chen, Hsuan-Cheng Huang, and
  Hsueh-Fen Juan.
\newblock Integrative network analysis reveals active micrornas and their
  functions in gastric cancer.
\newblock \emph{BMC systems biology}, 5\penalty0 (1):\penalty0 99, 2011.

\bibitem[Krek et~al.(2005)Krek, Gr{\"u}n, Poy, Wolf, Rosenberg, Epstein,
  MacMenamin, Da~Piedade, Gunsalus, Stoffel, et~al.]{krek2005combinatorial}
Azra Krek, Dominic Gr{\"u}n, Matthew~N Poy, Rachel Wolf, Lauren Rosenberg,
  Eric~J Epstein, Philip MacMenamin, Isabelle Da~Piedade, Kristin~C Gunsalus,
  Markus Stoffel, et~al.
\newblock Combinatorial microrna target predictions.
\newblock \emph{Nature genetics}, 37\penalty0 (5):\penalty0 495, 2005.

\bibitem[John et~al.(2004)John, Enright, Aravin, Tuschl, Sander, and
  Marks]{john2004human}
Bino John, Anton~J Enright, Alexei Aravin, Thomas Tuschl, Chris Sander, and
  Debora~S Marks.
\newblock Human microrna targets.
\newblock \emph{PLoS biology}, 2\penalty0 (11):\penalty0 e363, 2004.

\bibitem[Lewis et~al.(2005)Lewis, Burge, and Bartel]{lewis2005conserved}
Benjamin~P Lewis, Christopher~B Burge, and David~P Bartel.
\newblock Conserved seed pairing, often flanked by adenosines, indicates that
  thousands of human genes are microrna targets.
\newblock \emph{cell}, 120\penalty0 (1):\penalty0 15--20, 2005.

\bibitem[Griffiths-Jones et~al.(2006)Griffiths-Jones, Grocock, Van~Dongen,
  Bateman, and Enright]{griffiths2006mirbase}
Sam Griffiths-Jones, Russell~J Grocock, Stijn Van~Dongen, Alex Bateman, and
  Anton~J Enright.
\newblock mirbase: microrna sequences, targets and gene nomenclature.
\newblock \emph{Nucleic acids research}, 34\penalty0 (suppl\_1):\penalty0
  D140--D144, 2006.

\bibitem[Pi{\~n}ero et~al.(2016)Pi{\~n}ero, Bravo, Queralt-Rosinach,
  Guti{\'e}rrez-Sacrist{\'a}n, Deu-Pons, Centeno, Garc{\'\i}a-Garc{\'\i}a,
  Sanz, and Furlong]{pinero2016disgenet}
Janet Pi{\~n}ero, {\`A}lex Bravo, N{\'u}ria Queralt-Rosinach, Alba
  Guti{\'e}rrez-Sacrist{\'a}n, Jordi Deu-Pons, Emilio Centeno, Javier
  Garc{\'\i}a-Garc{\'\i}a, Ferran Sanz, and Laura~I Furlong.
\newblock Disgenet: a comprehensive platform integrating information on human
  disease-associated genes and variants.
\newblock \emph{Nucleic acids research}, page gkw943, 2016.

\bibitem[Abbott et~al.(2014)Abbott, Nyre, Abrahante, Ho, Isaksson~Vogel, and
  Starr]{abbott2014candidate}
Kenneth~L Abbott, Erik~T Nyre, Juan Abrahante, Yen-Yi Ho, Rachel
  Isaksson~Vogel, and Timothy~K Starr.
\newblock The candidate cancer gene database: a database of cancer driver genes
  from forward genetic screens in mice.
\newblock \emph{Nucleic acids research}, 43\penalty0 (D1):\penalty0 D844--D848,
  2014.

\bibitem[Yang et~al.(2012)Yang, Li, Jiang, Zhou, and Qu]{yang2012chipbase}
Jian-Hua Yang, Jun-Hao Li, Shan Jiang, Hui Zhou, and Liang-Hu Qu.
\newblock Chipbase: a database for decoding the transcriptional regulation of
  long non-coding rna and microrna genes from chip-seq data.
\newblock \emph{Nucleic acids research}, 41\penalty0 (D1):\penalty0 D177--D187,
  2012.

\bibitem[Chou et~al.(2015)Chou, Chang, Shrestha, Hsu, Lin, Lee, Yang, Hong,
  Wei, Tu, et~al.]{chou2015mirtarbase}
Chih-Hung Chou, Nai-Wen Chang, Sirjana Shrestha, Sheng-Da Hsu, Yu-Ling Lin,
  Wei-Hsiang Lee, Chi-Dung Yang, Hsiao-Chin Hong, Ting-Yen Wei, Siang-Jyun Tu,
  et~al.
\newblock mirtarbase 2016: updates to the experimentally validated mirna-target
  interactions database.
\newblock \emph{Nucleic acids research}, 44\penalty0 (D1):\penalty0 D239--D247,
  2015.

\bibitem[Shi et~al.(2013)Shi, Xu, Zhang, Xu, Li, Wang, Zhao, Jiang, Guo, and
  Li]{shi2013walking}
Hongbo Shi, Juan Xu, Guangde Zhang, Liangde Xu, Chunquan Li, Li~Wang, Zheng
  Zhao, Wei Jiang, Zheng Guo, and Xia Li.
\newblock Walking the interactome to identify human mirna-disease associations
  through the functional link between mirna targets and disease genes.
\newblock \emph{BMC systems biology}, 7\penalty0 (1):\penalty0 101, 2013.

\bibitem[Harvell et~al.(2013)Harvell, Kim, O’Brien, Tan, Borges, Schedin,
  Jacobsen, and Horwitz]{harvell2013genomic}
Djuana~ME Harvell, Jihye Kim, Jenean O’Brien, Aik-Choon Tan, Virginia~F
  Borges, Pepper Schedin, Britta~M Jacobsen, and Kathryn~B Horwitz.
\newblock Genomic signatures of pregnancy-associated breast cancer epithelia
  and stroma and their regulation by estrogens and progesterone.
\newblock \emph{Hormones and Cancer}, 4\penalty0 (3):\penalty0 140--153, 2013.

\bibitem[Barrett et~al.(2012)Barrett, Wilhite, Ledoux, Evangelista, Kim,
  Tomashevsky, Marshall, Phillippy, Sherman, Holko, et~al.]{barrett2012ncbi}
Tanya Barrett, Stephen~E Wilhite, Pierre Ledoux, Carlos Evangelista, Irene~F
  Kim, Maxim Tomashevsky, Kimberly~A Marshall, Katherine~H Phillippy, Patti~M
  Sherman, Michelle Holko, et~al.
\newblock Ncbi geo: archive for functional genomics data sets—update.
\newblock \emph{Nucleic acids research}, 41\penalty0 (D1):\penalty0 D991--D995,
  2012.

\bibitem[Langfelder and Horvath(2008)]{langfelder2008wgcna}
Peter Langfelder and Steve Horvath.
\newblock Wgcna: an r package for weighted correlation network analysis.
\newblock \emph{BMC bioinformatics}, 9\penalty0 (1):\penalty0 559, 2008.

\bibitem[Jiang et~al.(2008)Jiang, Wang, Hao, Juan, Teng, Zhang, Li, Wang, and
  Liu]{jiang2008mir2disease}
Qinghua Jiang, Yadong Wang, Yangyang Hao, Liran Juan, Mingxiang Teng, Xinjun
  Zhang, Meimei Li, Guohua Wang, and Yunlong Liu.
\newblock mir2disease: a manually curated database for microrna deregulation in
  human disease.
\newblock \emph{Nucleic acids research}, 37\penalty0 (suppl\_1):\penalty0
  D98--D104, 2008.

\bibitem[Li et~al.(2013)Li, Qiu, Tu, Geng, Yang, Jiang, and Cui]{li2013hmdd}
Yang Li, Chengxiang Qiu, Jian Tu, Bin Geng, Jichun Yang, Tianzi Jiang, and
  Qinghua Cui.
\newblock Hmdd v2. 0: a database for experimentally supported human microrna
  and disease associations.
\newblock \emph{Nucleic acids research}, 42\penalty0 (D1):\penalty0
  D1070--D1074, 2013.

\bibitem[Lee et~al.(2013)Lee, Kuo, Lin, Oyang, Huang, and
  Juan]{lee2013microrna}
Chia-Hsien Lee, Wen-Hong Kuo, Chen-Ching Lin, Yen-Jen Oyang, Hsuan-Cheng Huang,
  and Hsueh-Fen Juan.
\newblock Microrna-regulated protein-protein interaction networks and their
  functions in breast cancer.
\newblock \emph{International journal of molecular sciences}, 14\penalty0
  (6):\penalty0 11560--11606, 2013.

\bibitem[Met()]{MetaP}
Metap, a program to combine p values.
\newblock URL \url{http://igm.cumc.columbia.edu/MetaP/metap.php}.

\bibitem[Shyamasundar et~al.(2016)Shyamasundar, Lim, and
  Bay]{shyamasundar2016mir}
Sukanya Shyamasundar, Jia~Pei Lim, and Boon~Huat Bay.
\newblock mir-93 inhibits the invasive potential of triple-negative breast
  cancer cells in vitro via protein kinase wnk1.
\newblock \emph{International journal of oncology}, 49\penalty0 (6):\penalty0
  2629--2636, 2016.

\bibitem[Landman(2014)]{landman2014role}
Erin~O Landman.
\newblock The role of mir-526b in cox-2 mediated human breast cancer
  progression and induction of stem-like phenotype via ep4 receptor signaling.
\newblock 2014.

\bibitem[Hossain et~al.(2006)Hossain, Kuo, and Saunders]{hossain2006mir}
Anwar Hossain, Macus~T Kuo, and Grady~F Saunders.
\newblock Mir-17-5p regulates breast cancer cell proliferation by inhibiting
  translation of aib1 mrna.
\newblock \emph{Molecular and cellular biology}, 26\penalty0 (21):\penalty0
  8191--8201, 2006.

\bibitem[Li et~al.(2012)Li, Zhang, Zhang, Jia, Kang, and
  Zhu]{li2012differential}
Jian-Yi Li, Yang Zhang, Wen-Hai Zhang, Shi Jia, Ye~Kang, and Xiao-Yu Zhu.
\newblock Differential distribution of mir-20a and mir-20b may underly
  metastatic heterogeneity of breast cancers.
\newblock \emph{Asian Pacific Journal of Cancer Prevention}, 13\penalty0
  (5):\penalty0 1901--1906, 2012.

\bibitem[Chen et~al.(2016)Chen, Shin, Siu, Ho, Cheuk, and Kwong]{chen2016mir}
Jiawei Chen, Vivian~Y Shin, Man~T Siu, John~CW Ho, Isabella Cheuk, and Ava
  Kwong.
\newblock mir-199a-5p confers tumor-suppressive role in triple-negative breast
  cancer.
\newblock \emph{BMC cancer}, 16\penalty0 (1):\penalty0 887, 2016.

\bibitem[Li et~al.(2015)Li, Shan, Chen, Zhang, Su, Huang, Yu, Zhi, Li, Wang,
  et~al.]{li2015epigenetic}
Pu~Li, Jing-Xuan Shan, Xue-Hua Chen, Di~Zhang, Li-Ping Su, Xiu-Ying Huang,
  Bei-Qin Yu, Qiao-Ming Zhi, Cheng-Long Li, Ya-Qing Wang, et~al.
\newblock Epigenetic silencing of microrna-149 in cancer-associated fibroblasts
  mediates prostaglandin e2/interleukin-6 signaling in the tumor
  microenvironment.
\newblock \emph{Cell research}, 25\penalty0 (5):\penalty0 588, 2015.

\bibitem[Fang et~al.(2016)Fang, Wang, Li, and Zeng]{fang2016down}
Cheng Fang, Fu-Bing Wang, Yirong Li, and Xian-Tao Zeng.
\newblock Down-regulation of mir-199b-5p is correlated with poor prognosis for
  breast cancer patients.
\newblock \emph{Biomedicine \& Pharmacotherapy}, 84:\penalty0 1189--1193, 2016.

\bibitem[Lehmann et~al.(2010)Lehmann, Streichert, Otto, Albat, Hasemeier,
  Christgen, Schipper, Hille, Kreipe, and
  L{\"a}nger]{lehmann2010identification}
Ulrich Lehmann, Thomas Streichert, Benjamin Otto, Cord Albat, Britta Hasemeier,
  Henriette Christgen, Elisa Schipper, Ursula Hille, Hans~H Kreipe, and Florian
  L{\"a}nger.
\newblock Identification of differentially expressed micrornas in human male
  breast cancer.
\newblock \emph{BMC cancer}, 10\penalty0 (1):\penalty0 109, 2010.

\bibitem[Cheng et~al.(2012)Cheng, Wang, Chang, Chu, Chen, Yu, Chao, Liu, Ding,
  and Shen]{cheng2012microrna}
Chun-Wen Cheng, Hsiao-Wei Wang, Chia-Wei Chang, Hou-Wei Chu, Cheng-You Chen,
  Jyh-Cherng Yu, Jui-I Chao, Huei-Fang Liu, Shian-ling Ding, and Chen-Yang
  Shen.
\newblock Microrna-30a inhibits cell migration and invasion by downregulating
  vimentin expression and is a potential prognostic marker in breast cancer.
\newblock \emph{Breast cancer research and treatment}, 134\penalty0
  (3):\penalty0 1081--1093, 2012.

\end{thebibliography}

\end{document}